\documentclass[useAMS,usenatbib]{mnras}
\usepackage{epsfig}
\usepackage{graphicx}
\usepackage{amsfonts}
\usepackage[usenames,dvipsnames]{xcolor}
\usepackage{url}
\usepackage{subfigure}
\usepackage{times,graphicx,amsmath,amsfonts,amssymb,aas_macros,epstopdf,hyperref}
\usepackage[normalem]{ulem}
\usepackage[T1]{fontenc}
\usepackage{multirow}
\usepackage{mathrsfs}
\usepackage{ae,aecompl}
\usepackage{xspace} 
\usepackage{units}
\usepackage{wasysym}
\usepackage{newtxtext,newtxmath}

\hypersetup{
    colorlinks=true,
    linkcolor=blue,
    citecolor=blue,
    urlcolor=blue,
}

\def\etal{{\frenchspacing\it et al.}}
\def\ie{{\frenchspacing\it i.e.}}

\def\be{\begin{equation}}
\def\ee{\end{equation}}
\def\ba{\begin{eqnarray}}
\def\ea{\end{eqnarray}}

\newcommand{\hompc}{\,h\,{\rm Mpc}^{-1}}
\newcommand{\mpcoh}{\,h^{-1}\,{\rm Mpc}}
\newcommand{\pkunit}{\,h^{-3}\,{\rm Mpc}^3}

\newcommand{\ab}{\alpha_{\bot}}
\newcommand{\ap}{\alpha_{\|}}
\newcommand{\az}{\alpha_{0}}
\newcommand{\aone}{\alpha_{1}}

\newcommand{\remove}[1]{}

\def\LaTeX{L\kern-.36em\raise.3ex\hbox{a}\kern-.15em
    T\kern-.1667em\lower.7ex\hbox{E}\kern-.125emX}
    
\title[An anisotropic BAO analysis of the eBOSS DR14 QSO sample]{The clustering of the SDSS-IV extended Baryon Oscillation Spectroscopic Survey DR14 quasar sample: Anisotropic Baryon Acoustic Oscillations measurements in Fourier-space with optimal redshift weights}
\author[Wang, Zhao \& Wang \etal]
{\parbox{\textwidth}{Dandan Wang$^{1,2}$\thanks{E-mail: \url{ddwang@nao.cas.cn}}, Gong-Bo Zhao$^{1,3,4}$\thanks{E-mail: \url{gbzhao@nao.cas.cn}}, Yuting Wang$^{1}$, Will J. Percival$^{4}$, Rossana Ruggeri$^{4}$, Fangzhou Zhu$^{5}$, Rita Tojeiro$^{6}$, Adam D. Myers$^{7}$, Chia-Hsun Chuang$^{8,9}$, Falk Baumgarten$^{8,10}$, Cheng Zhao$^{11}$, H\' ector Gil-Mar\' in$^{12,13}$, Ashley J. Ross$^{14,4}$, Etienne Burtin$^{15}$, Pauline Zarrouk$^{15}$, Julian Bautista$^{4}$, Jonathan Brinkmann$^{16}$, Kyle Dawson$^{17}$, Joel R. Brownstein$^{17}$, Axel de la Macorra$^{18}$, Donald P. Schneider$^{19,20}$, Arman Shafieloo$^{21,22}$} \vspace*{20pt} \\
\parbox{\textwidth}{
$^{1}$ National Astronomical Observatories, Chinese Academy of Science, Beijing 100012, P.R.China\\
$^2$ University of Chinese Academy of Sciences, Beijing 100049, P.R.China\\
$^3$ College of Astronomy and Space Sciences, University of Chinese Academy of Sciences, Beijing 100049, P.R.China\\
$^{4}$Institute of Cosmology \& Gravitation, Dennis Sciama Building, University of Portsmouth, Portsmouth, PO1 3FX, UK\\
$^{5}$Department of Physics, Yale University, 260 Whitney Ave, New Haven, CT 06520, USA\\
$^{6}$School of Physics and Astronomy, University of St Andrews, North Haugh, St Andrews KY16 9SS, UK\\
$^{7}$ Department of Physics and Astronomy, University of Wyoming, Laramie, WY 82071, USA\\
$^{8}$Leibniz-Institut f\"ur Astrophysik Potsdam (AIP), An der Sternwarte 16, D-14482 Potsdam, Germany\\
$^{9}$Kavli Institute for Particle Astrophysics and Cosmology \& Physics Department, Stanford University, Stanford, CA 94305, USA\\
$^{10}$Humboldt-Universit\"at zu Berlin, Institut f\"ur Physik, Newtonstrasse 15, D-12589 Berlin, Germany\\
$^{11}$Tsinghua Center for Astrophysics and Department of Physics, Tsinghua University, Beijing 100084, China\\
$^{12}$Sorbonne Universit\' es, Institut Lagrange de Paris (ILP), 98 bis Boulevard Arago, 75014 Paris, France\\
$^{13}$Laboratoire de Physique Nucl\' eaire et de Hautes Energies, Universit\' e Pierre et Marie Curie, 4 Place Jussieu, 75005 Paris, France\\
$^{14}$Center for Cosmology and Astro-Particle Physics, Ohio State University, Columbus, Ohio, USA\\
$^{15}$IRFU,CEA, Universit\' e Paris-Saclay, F-91191 Gif-sur-Yvette, France\\
$^{16}$Apache Point Observatory, P.O. Box 59, Sunspot, NM 88349\\
$^{17}$Department Physics and Astronomy, University of Utah, 115 S 1400 E, Salt Lake City, UT 84112, USA\\
$^{18}$Instituto de F\' isica, Universidad Nacional Aut\' onoma de México, Apdo.Postal 20-364, 01000, M\' exico D.F., M\' exico\\
$^{19}$Department of Astronomy and Astrophysics, The Pennsylvania State University, University Park, PA 16802, USA\\
$^{20}$Institute for Gravitation and the Cosmos, The Pennsylvania State University, University Park, PA 16802, USA\\
$^{21}$Korea Astronomy and Space Science Institute, Yuseong-gu, 776 Daedeok daero, Daejeon 34055, Korea\\
$^{22}$University of Science and Technology, Yuseong-gu 217 Gajeong-ro, Daejeon 34113, Korea\\
}}
\begin{document}

\label{firstpage}
\maketitle

\begin{abstract}
 
We present a measurement of the anisotropic and isotropic Baryon Acoustic Oscillations (BAO) from the extended Baryon Oscillation Spectroscopic Survey Data Release 14 quasar sample with optimal redshift weights. Applying the redshift weights improves the constraint on the BAO dilation parameter $\alpha(z_{\rm eff})$ by 17\%. We reconstruct the evolution history of the BAO distance indicators in the redshift range of $0.8<z<2.2$. This paper is part of a set that analyses the eBOSS DR14 quasar sample.

\end{abstract}


\section{Introduction}
\label{sec:intro}

Unveiling the underlying physics of the accelerating expansion of the universe has been one of the most challenging tasks in cosmology since its discovery from observations of supernovae \citep{Riess,Perlmutter}. Largely complementary to the supernovae, the Baryon Acoustic Oscillations (BAO), as a `cosmic stander ruler', has become one of the most robust cosmological probes of the expansion history of the Universe since it was first detected \citep{2005MNRAS.362..505C,2005ApJ...633..560E} from large scale galaxy surveys.

To map the evolution of the cosmic expansion, which is crucial to study the nature of dark energy, BAO measurements at various cosmic epochs are required \citep{ZhaoDE12, ZhaoDE17}. However, it is challenging to extract the tomographic BAO information from a galaxy survey, as usually one has to combine galaxies from a range of redshifts to obtain a robust BAO measurement at one, or a small number of {\it effective} redshifts. For example, the SDSS-III BOSS \citep{Eisenstein11,Dawson12} survey has successfully obtained a per cent level accuracy BAO measurement, but only at three effective redshifts in the range of $0.2<z<0.75$ \citep{Alam16}.

To extract the lightcone information, one can decompose the survey into a large number of overlapping redshift slices, and perform the BAO analysis in each redshift bin, and every pair of redshift bins to quantify the covariance \citep{Zhaotomo16,Wangtomo16}. However, this approach is computationally expensive, and is likely to be impractical for future deep surveys such as DESI or Euclid.

A more efficient approach is to assign each galaxy an additional weight, according to its redshift, and optimise the weight so that a high level of tomographic information can be extracted at a low computational cost. Early applications of the optimal redshift weight for a BAO measurement were made by \cite{Zhu:2014ica,zwBAOmock} in configuration space using two-point correlation functions. In this work, we adopt a complementary approach to perform a new BAO analysis with optimal redshift weights in Fourier space, and apply our technique to the extended Baryon Oscillation Spectroscopic Survey (eBOSS) \citep{Dawson:2015wdb} Data Release 14 (DR14) quasar (QSO) sample. 

The structure of this paper is as follows. We introduce the DR14 quasar catalogue and mocks used in this work in Section 2; then in Section 3, we describe details of the methodology, including the derivation of the redshift weights for power spectrum multipoles, the measurement of the weighted sample, the template, and fitting procedure used for this analysis. We present our main results in Section 4, followed by a conclusion and discussion in Section 5. 

\section{THE DR14 QSO and mock samples}

eBOSS \citep{KD16} is a cosmological survey of the SDSS-IV project \citep{Blanton2017}, which uses the 2.5-meter Sloan Telescope \citep{Gunn:2006tw} with the BOSS double-armed spectrographs \citep{2013AJ....146...32S}. The eBOSS DR14 quasar catalogue \citep{Paris2017,DR14} that we use contains quasars from the first two years of eBOSS observations, which are limited to a redshift range of $0.8<z<2.2$. The catalogue covers 1214.64 $\rm deg^2$ in the North Galactic Cap (NGC), and 898.27 $\rm deg^2$ in the South Galactic Cap (SGC). There are 95161 and 63596 effective quasars in the NGC and SGC, respectively. The effective volume is $5.44\times 10^7 (\mpcoh) ^3$ and $3.34\times 10^7 (\mpcoh) ^3$ for the NGC and SGC, respectively. The redshifts are adopted from the SDSS quasar pipeline ($\rm Z_{\rm{PL}}$) with visual inspections. 

\begin{figure}
\centering
{\includegraphics[width=0.5\textwidth]{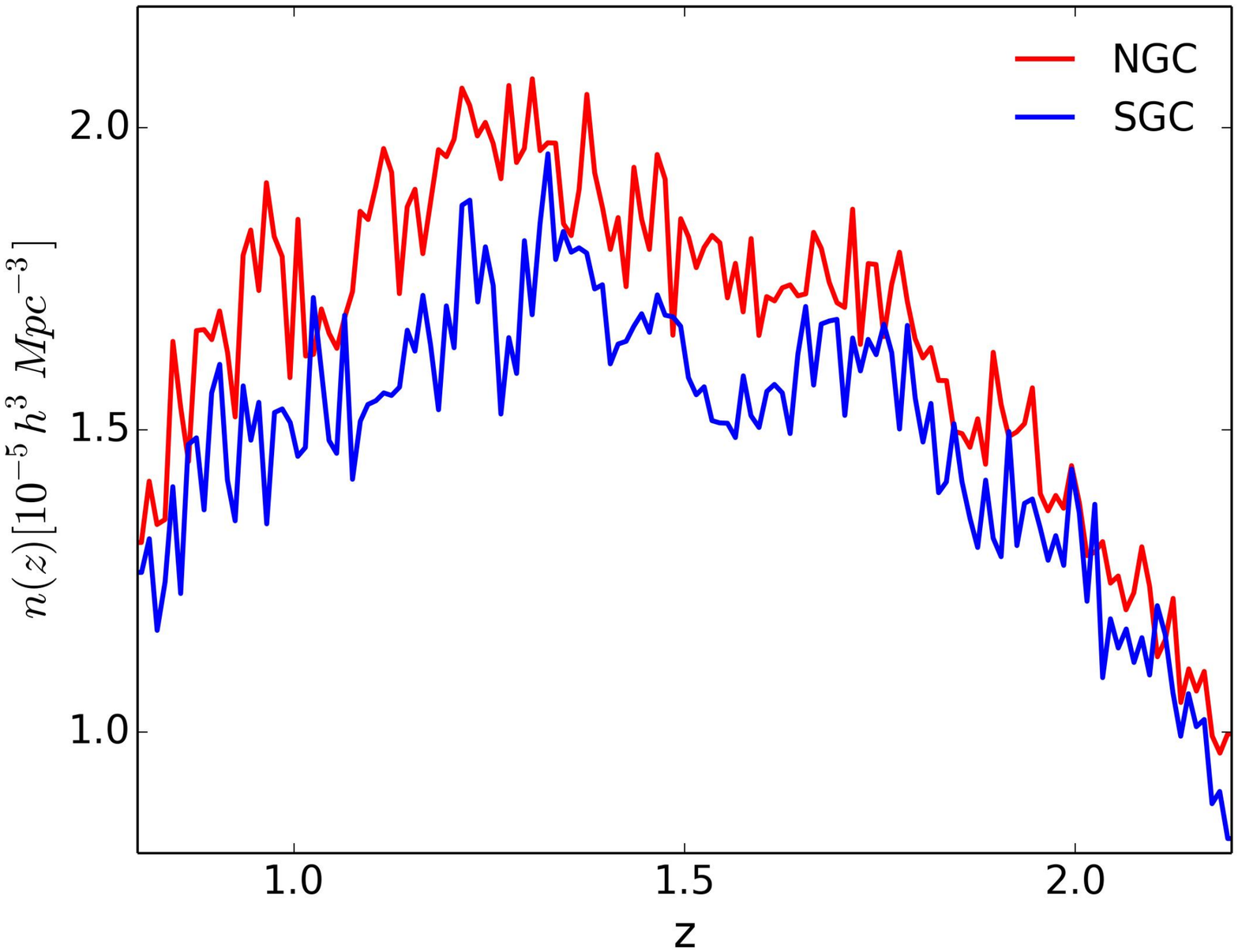}}
\caption{The redshift distribution for the DR14 QSO sample in the NGC (red) and SGC (blue).}
\label{fig:nbar}
\end{figure}

Fig \ref{fig:nbar} shows the redshift distribution of the quasar sample. There is a slight difference in the NGC and the SGC, because the targeting efficiency in these two regions is slightly different. More details of the target selection can be found in \cite{Myers2015}.

Each quasar in the DR14 quasar catalogue is assigned a product of a few different weights, namely, $w_{\rm sys}$, $w_{\rm cp}$, $w_{\rm focal}$ and  $w_{\rm FKP}$, where $w_{\rm sys}$ is the systematic weight correcting for effects such as Galactic extinction and the limiting magnitude; the close pair weight $w_{\rm cp}$ and the focal plane weight $w_{\rm focal}$ correct for redshift failures and fibre collisions, and the FKP weight $w_{\rm FKP}$ minimises the uncertainty of power spectrum measurement as introduced by \cite{1994ApJ...426...23F},
\be w_{\rm FKP} = \frac{1}{1+\bar{n}(z)P_{0}} \ee where
$P_{0}$ is the amplitude of power spectrum in $k$ space, which is fixed to $6000 \ \pkunit$ in this paper.
In addition, we assign each quasar a redshift weight $w_{z}$, which is detailed in Section \ref{sec:weight}. Thus the total weight for each quasar is
\be w_{\rm tot}=w_{\rm sys}w_{\rm cp} w_{\rm focal} w_{\rm FKP}  \sqrt{w_{z}} \ .\ee

\begin{table}
\begin{center}
\begin{tabular}{c c c c c}
\hline\hline
  & $\Omega_{m}$ & $\Omega_{\Lambda}$ & $\Omega_{b}h^{2}$& $h$\\ \hline
\hline
Fiducial  & $0.31 $ & $0.69$ & $0.022$& $0.676$ \\
EZmock & $0.307$ & $0.693$ & $0.02214$ &$0.676$\\ 
\hline
\end{tabular}
\caption{The fiducial cosmological model of this paper, the cosmology used in creating the EZ mocks.}
\label{tabmodel}
\end{center}
\end{table}

We use 1000 EZmocks \citep{Chuang:2015} to compute the data covariance matrix and for mock tests. EZmocks has the light-cone information, which allows one to investigate the redshift evolution of the clustering of quasars. The fiducial cosmology used for EZmocks is given in Table \ref{tabmodel}. 

\section{Methodology}

\subsection{The optimal redshift weights}
\label{sec:weight}

In this section, we present details of the algorithm of optimal redshift weighting for BAO analysis in Fourier space.

We use the parametrisation of the distance-redshift relation described in \cite{Zhu:2014ica},
\be\label{eq:chi} \frac{\chi(z)}{\chi_{\rm{fid}}(z)}=\alpha_{0}\left(1+\alpha_{1} x+\frac{1}{2} \alpha_{2} x^2 +\cdots \right) \ee
where $x=\chi_{\rm{fid}}(z) / \chi_{\rm{fid}}(z_{0})-1$, the subscript `fid' denotes the fiducial cosmology (Table \ref{tabmodel}). As demonstrated in \cite{doi:stw1515}, Eq (\ref{eq:chi}) can accurately parametrise $\chi(z)$ for a wide range of cosmologies. In this work, we set the pivot redshift $z_0$ to be the effective redshift of the quasar sample, \ie, $z_0=1.52$.

The transverse and the radial BAO dilation parameters $\alpha_{\perp}$ and $\alpha_{\parallel}$ are, 
\ba \label{eq:az} && \alpha_{\perp}= \alpha_{0}\left(1+\alpha_{1} x+\frac{1}{2}\alpha_{2}x^2 \right)  \nonumber \\
 && \alpha_{\parallel} = \alpha_{0}\left(1+\alpha_{1}+\left(2\alpha_{1}+\alpha_{2} \right)x+\frac{3}{2}\alpha_{2}x^2\right)  \ea

The optimal redshift weight of $\alpha_{i}$ can be evaluated as follows \citep{Zhu:2014ica},
\be \boldsymbol{w_{\ell,i}} = \boldsymbol{C^{-1}} \boldsymbol{P_{\ell,i}} \nonumber \ee
where $\ell$ refers to the power spectrum multipole, and $\boldsymbol{C}$ is the data covariance matrix,
\be \boldsymbol{C} = \left( P + \frac{1}{\bar n} \right) ^2 dV \nonumber \ee

The diagonal elements of $\boldsymbol{C^{-1}}$ essentially represent the effective volume of the survey at various redshifts. As the light-cone of EZmocks is assembled by snapshots at seven redshifts, we split the entire redshift range into seven slices, and compute $\bar n(z)$ in each redshift slice.

The quantity $P_{\ell,i}$ is the derivative of the power spectrum multipole with respect to $\alpha_i$, which can be evaluated analytically, \be \frac{\partial P_{\ell}(k,z)}{\partial\alpha_{i}} = \frac{\partial P_{\ell}(k,z)}{\partial\alpha_{\parallel}} \frac{\partial\alpha_{\parallel}}{\partial\alpha_{i}}+\frac{\partial P_{\ell}(k,z)}{\partial\alpha_{\perp}} \frac{\partial\alpha_{\perp}}{\partial\alpha_{i}} \nonumber \ee where $P_{\ell}(k,z)$ is the $l^{\rm th}$ power spectrum multipole at wavenumber $k$ and redshift $z$, as detailed in Section \ref{sec:temp}.

Given Eq (\ref{eq:az}), it is straightforward to obtain several of the derivative terms analytically,
\be \frac{\partial\alpha_{\perp}}{\partial\alpha_{0}}=1;\qquad            \frac{\partial\alpha_{\parallel}}{\partial\alpha_{0}}=1 \nonumber \ee
\be \frac{\partial\alpha_{\perp}}{\partial\alpha_{1}}=x;\qquad
\frac{\partial\alpha_{\parallel}}{\partial\alpha_{1}}=1+2x \nonumber \ee
\be \label{eq:functionx} \frac{\partial\alpha_{\perp}}{\partial\alpha_{2}}=\frac{1}{2}x^2;\qquad
\frac{\partial\alpha_{\parallel}}{\partial\alpha_{2}}=x+\frac{3}{2}x^2  \ee 
and the terms $\frac{\partial P_{\ell}(k,z)}{\partial\ab}$ and $\frac{\partial P_{\ell}(k,z)}{\partial\ap}$ can be evaluated numerically\footnote{These terms can also be evaluated analytically with approximations \citep{zwRSD}.}. 

\begin{figure}
\centering
{\includegraphics[width=0.5\textwidth]{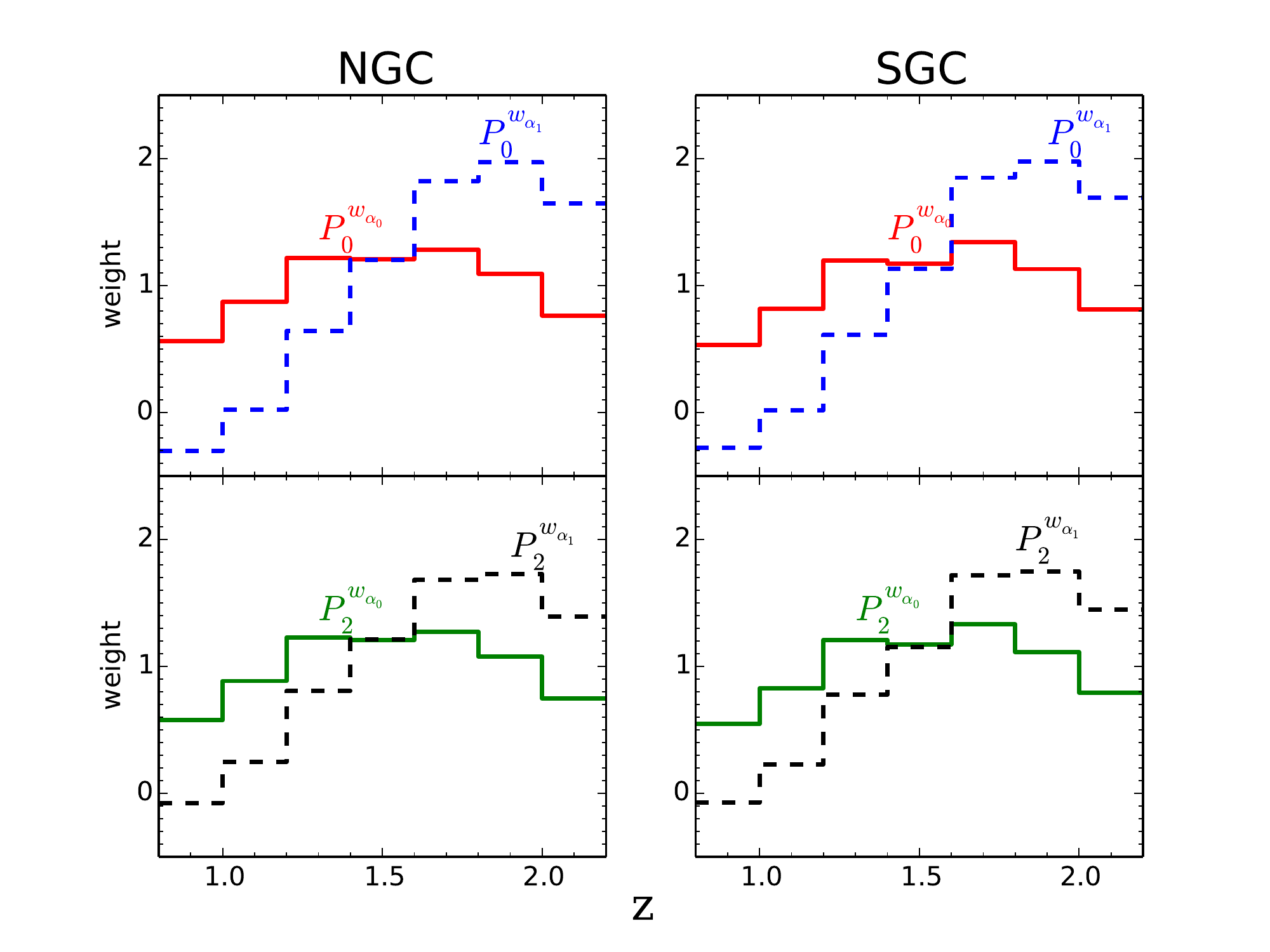}}
\caption{The optimal redshift weights derived from power spectra multipoles measured in NGC (left) and SGC (right), respectively. In all panels, weights for $\alpha_{0}$ and $\alpha_{1}$ are shown in solid and dashed lines respectively. 
}
\label{fig:weight}
\end{figure}

\begin{figure*}
\centering
{\includegraphics[scale=0.5]{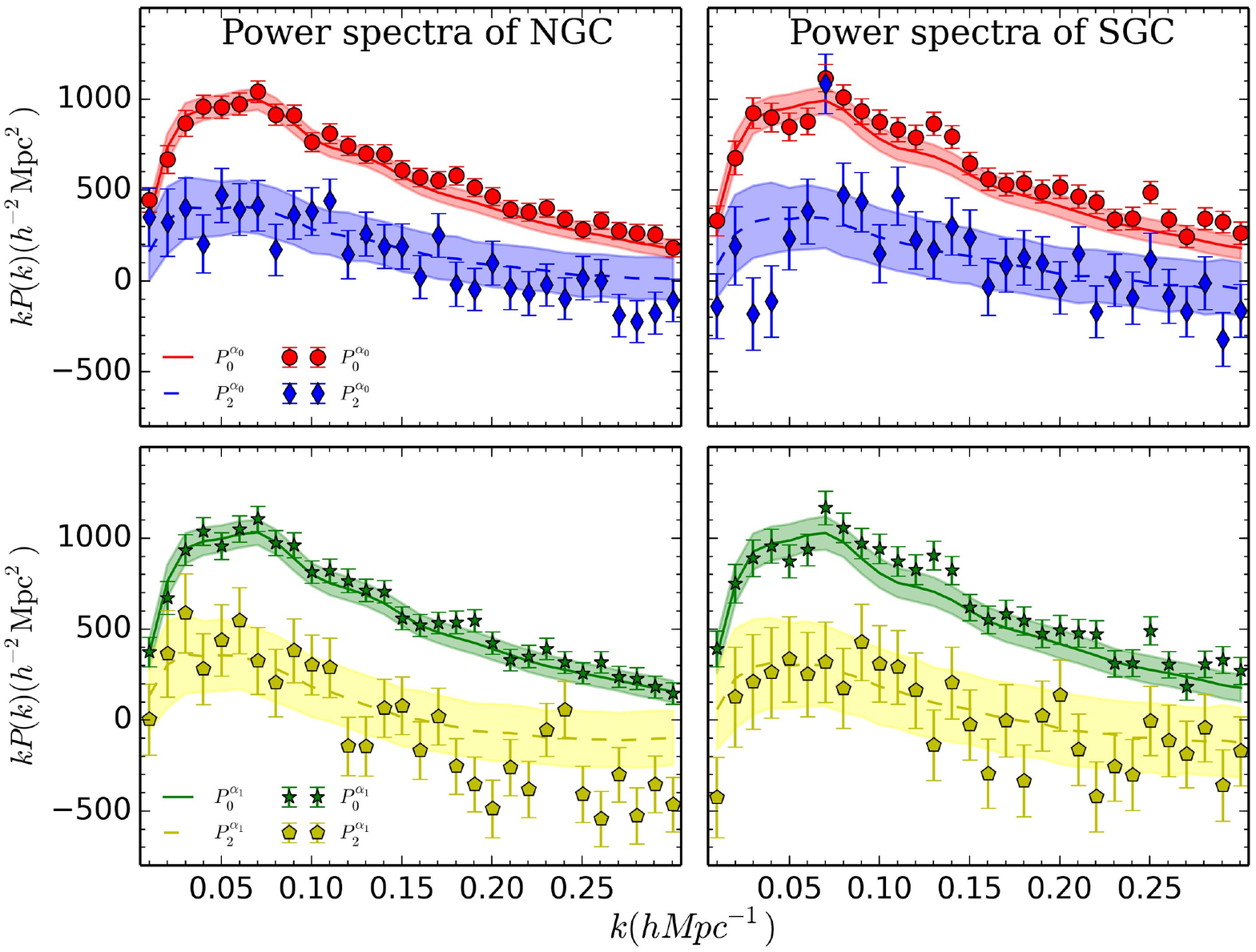}}
\caption{The power spectrum monopole and quadrupole measured from the DR14 quasar catalogue (data points with error bars) and from 1000 EZmocks (shaded bands) weighted by the optimal redshift weights for $\alpha_0$ (upper panels) and $\alpha_1$ (lower) in the NGC (left panels) and SGC (right) respectively. 
}
\label{fig:pk}
\end{figure*}

\begin{figure*}
\centering
{\includegraphics[width=0.75\textwidth]{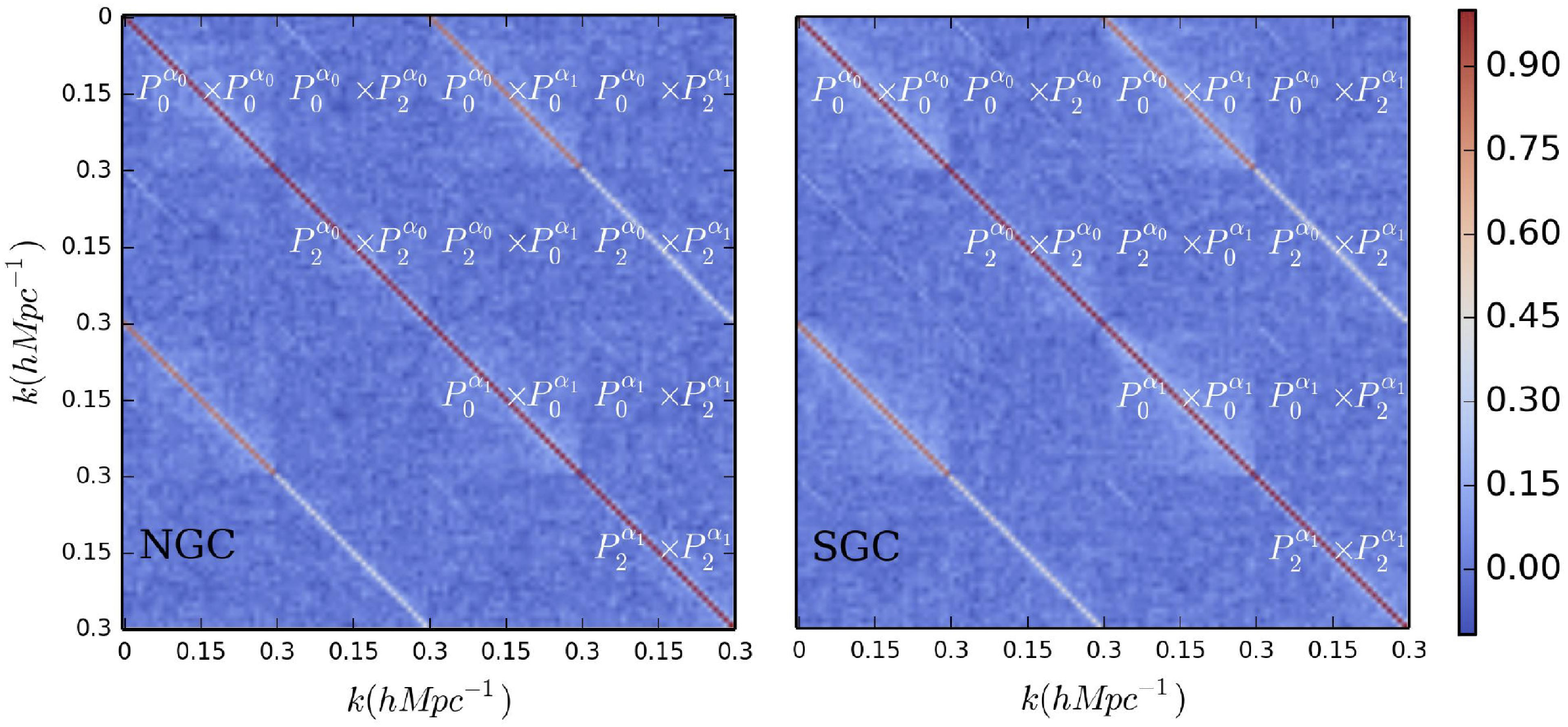}}
\caption{The correlation matrix among power spectra monopole and quadrupole weighted by $\az$ and $\aone$ in the NGC (left) and SGC (right), respectively. 
}
\label{fig:cor}
\end{figure*}

These weights are generally functions of $z$ and $k$, we have numerically checked that the $k$-dependence of the weights and the dependence is so weak in the $k$ range of interest that we drop the $k$-dependence for simplicity, and evaluate the weights at $k=0.1 \ \hompc$.

The resultant redshift weights are shown in Fig \ref{fig:weight} \footnote{We derive the weights at the resolution of $\Delta z=0.2$ as this bin size is the redshift resolution used to generate the lightcone for the EZ mocks.}. The weights for the monopole and quadrupole are similar, and they all peak at the effective redshift. This behavior is expected as $P_0$ and $P_2$ evolve with redshift in similar ways \footnote{$P_{\rm 0}$ and $P_{\rm 2}$ evovle in exactly the same way in linear perturbation theory.}, and the monopole is most sensitive to an isotropic BAO shift parametrised by $\az$ at the effective redshift, where the monopole has the largest signal to noise ratio. On the other hand, power spectra at high redshifts are more useful to measure $\aone$, as it is apparent from Eq (\ref{eq:az}) that the effect of $\aone$ on the BAO measurement is maximised at high $z$. This is the reason why the weights for $\az$ peaks at high redshifts.

\remove{ shows the redshift weighting it illustrates that monopole weighting of $\alpha_{0}$ and quadrupole weighting of $\alpha_{0}$ are quite similar, but the corresponding weightings of $\alpha_{1}$ are different. That is because power spectrum multipoles are ,theoretically, all in similar shape, and $\alpha_{0}$ just rescale power spectrum multipoles by a constant factor at any redshift. However, $\alpha_{1}$ affects the shape of power spectrum multipoles differently at different redshift, after combining all power spectrum multipoles together, the shape change more obvious. But the reshape is not prominent severely, the gap of the weights are not significant. There is still slight difference between weights for the NGC and the SGC, due to different distribution of quasar samples. }

\subsection{Measurements of the power spectra multipoles}

\remove{The redshift weightings at redshift z are corresponding to the auto-correlation pairs at z, but it is quite convenient to apply the square root of weightings to individual galaxy. However, as Fig \ref{fig:weight} shows, we cannot avoid that the weighting of $\alpha_{1}$ in the first redshift bin is negative. We use two set of coefficients to linearly combine weighting of $\alpha_{0}$ and $\alpha_{1}$, the combined weightings are positive, only need make sure the Jacob matrix of the coefficient matrix is not 0. (Rossana et al perp). We provide the combined weightings for real measurements, and the decompose the power spectrum multipoles proportionally.}

To apply the redshift weights to the quasar and random catalogues, we first perform a linear transformation of the weights to get a set of positive-definite new weights, which is required as the weight assigned to each quasar is the square root of the $z$-weights derived previously. As this is a linear operation, this transformation preserves the information content.

To measure the power spectrum multipoles from the $z$-weighted DR14 quasar catalogue and each of the EZ mock catalogues, we adopt the method detailed in \cite{Zhaotomo16}, which is based on a Fast Fourier Tansform (FFT) method \citep{PkFFT}. We embed the entire survey volume into a cubic box with size of 8000 $\mpcoh$ a side, and subdivide it into $N_{\rm g}=1024^3$ grids. We use the Piecewise Cubic Spline (PCS) interpolation to smooth the overdensity field to reduce the aliasing effect when assigning the quasar samples and randoms to the grids, \cite{PCS}.

We measure the multipoles up to $k=0.3 \hompc$ with $\Delta k =0.01\hompc$. Fig. \ref{fig:pk} displays the power spectrum monopole and quadrupole weighted for $\az$ and $\aone$ in both NGC and SGC. The observables in the NGC and SGC are consistent within the error bars derived from the EZ mocks. 

The data covariance matrix is computed as,

\ba
 C_{ij,\alpha_{m} \alpha_{n}}^{\ell \ell^{\prime}} &=&\frac{1}{N_{\rm mock}-1} \sum_{q=1} ^{N_{\rm{mock}}}\left[P_{\ell,\alpha_{m}} ^{q} (k_{i})- \bar{P}_{\ell,\alpha_{m}}(k_{i})\right] \nonumber \\
 && \times \ \left[P_{\ell^{\prime},\alpha_{n}} ^ {q} (k_{j})- \bar{P}_{\ell^{\prime},\alpha_{n}}(k_{j})\right]
\label{covariance} \ea 
 \be \bar{P}_{\ell,\alpha_{m}} (k_{i}) = \frac{1}{N_{\rm{mock}}} \sum_{q=1} ^{N_{\rm{mock}}} P_{\ell,\alpha_{m}}^{q}(k_{i}) \ee where $N_{\rm mock}=1000$, $i$ denotes the $i^{\rm th}$ $k$-bin, $\ell$ denotes order of the power spectrum multipole, and $m$ runs over ${0,1}$.  
 
We use the Hartlap factor $f_{\rm H}$ to correct for the bias of the inverse of the maximum-likelihood estimator of the covariance matrix \citep{Hartlap:2006kj}. The factor $f_{\rm H}$ is defined as,
\be f_{\rm H}=\frac{N_{\rm{mock}}-N_{\rm b}-2}{N_{\rm{mock}}-1} \nonumber \ee
where $N_{\rm b}$ is the number of $k$-bins used for analysis. The corrected inverse matrix of the covariance matrix is,
\be \widetilde{C}_{ij} ^{-1}=f_{\rm H}C_{ij} ^{-1} \nonumber \ee
 
The corresponding data correlation matrices for these observables, which are the normalised data covariance matrices with all the diagonal elements being unity, for these observation, are presented in Fig \ref{fig:cor}. The $\az$ and $\aone$ weighted monopoles correlate with each other (at the same $k$ mode) to a large extent, thus it is difficult to constrain $\az$ and $\aone$ simultaneously using  the monopole alone. However, the correlation for the quadrupole is much less; adding quadrupole to the analysis can assist in breaking the degeneracy between $\az$ and $\aone$.

\subsection{The BAO analysis}
\label{analysis}

\subsubsection{The template}
\label{sec:temp}

The template we chose to model the $z$-dependent two-dimensional quasar power spectrum is,

\ba
P_{g}(k, \mu, z) &= &P_{\rm nw,lin} (k, \mu, z) \left[b(z)+f(z) \mu ^2 \right]^2  \nonumber \\
&&\left[1+ O_{\rm lin}(k)e^{-k^2 ( \mu ^2 \Sigma_{\parallel} ^2 +(1-\mu ^2) \Sigma_{\perp} ^2 )/2}\right]
\ea
\be P_{\rm{nw,lin}} (k,z)= \left[ \frac{D(z)}{D(z=z_{0})} \right] ^2 P_{\rm nw,lin}(k,z_{0})  \ee
where $D(z)$ is the growth function. We follow \cite{Ata:2017dya} and fix $\Sigma_{\perp}$  to $7.8\mpcoh$ and $\Sigma_{\parallel}$ to $5.2\mpcoh$ at the effective redshift.

We model the time evolution of the linear bias using the quadratic function of $b(z)=0.53+0.29(1+z)^2$ \citep{croom}, which has been confirmed to be a reasonable model for the eBOSS DR14 sample \citep{QSObias}. The linear growth rate $f(z)$ is modelled follows \cite{Lindergamma},
\be f(z)=\left[ \frac{\Omega_{m} (1+z)^3}{\Omega_{m}(1+z)^3 +1-\Omega_{m}} \right] ^{\gamma} \nonumber \ee
where the gravitational growth index $\gamma$ is fixed to 0.545.

The multipole can be calculated from,
\ba P_{\ell}(k,z)&=& \frac{2\ell+1}{2\alpha_{\perp}^2 \alpha_{\parallel}} \int  P_{g}(k^{\prime},\mu ^{\prime},z) \mathcal{L}(\mu) d\mu \label{kdepend} \\ 
&&+\frac{a_{\ell 1}}{k^3}+\frac{a_{\ell 2}}{k^2}+\frac{a_{\ell 3}}{k}+a_{\ell 4}+a_{\ell 5}k \ea with
\ba k^{\prime} = \frac{k}{\alpha_{\perp} \sqrt{1+ \left[ \left(\frac{\alpha_{\perp}}{\alpha_{\parallel}}\right)^2 -1\right] \mu ^2}} \nonumber \\
 \mu ^{\prime} =\frac{\mu}{\frac{\alpha_{\parallel}}{\alpha_{\perp}} \sqrt{1+\left[ \left(\frac{\alpha_{\perp}}{\alpha_{\parallel}}\right)^2 -1\right] \mu ^2}} \label{11} \ea to encode the Alcock-Paczynski effect \citep{APeffect}, where
\be \alpha_{\perp} = \frac{D_{A}(z)r_{\rm d}^{\rm fid}}{D_{A}^{\rm fid}(z) r_{\rm d}};\qquad      \alpha_{\parallel}=\frac{H^{\rm fid}(z) r_{\rm d}^{\rm fid}}{H(z)r_{\rm d}} \nonumber \ee and the polynomial is to marginalised over the broad band shape.

The $z$-weighted template is,
\be  P_{\ell}^{W_i}(k) = \int w_{\ell,\alpha_{i}}(z) P_{\ell}(k,z) dz \ee

\subsubsection{Parameter estimation}

We perform the parameter estimation using a modified version of {\tt CosmoMC}, which is a Markov Chain Monte Carlo (MCMC) \citep{Lewis:2002ah} engine for cosmological parameter constraints. We minimise the following $\chi^2$ in the global fitting, \be \chi^2 = (D+X)^T C^{-1} (D+X) \nonumber \ee where $D$ is the difference vector defined as $D(k)\equiv P^{\rm data}(k)-P^{\rm theo}(k)$, and $C$ is the covariance matrix for the weighted power spectra multipoles. We follow the method presented in \cite{Zhao:2016das} to analytically marginalise over the nuisance parameters $a_{\ell i}$. We correct for the bias due to our finite number of mocks by rescaling the inverse data covariance matrix by the $M$ factor \citep{Percival:2013sga}, 
\be M=\sqrt{\frac{1+B(N_{\rm b}-N_{\rm p})}{1+A+B(N_{\rm p}+1)}} \nonumber \ee
with
\be A=\frac{2}{(N_{\rm{mock}}-N_{\rm b}-1)(N-N_{\rm b}-4)} \nonumber \ee
\be B=\frac{N_{\rm{mock}}-N_{b}-2}{(N_{\rm mock}-N_{\rm b}-1)(N_{\rm{mock}}-N_{\rm b}-4)} \nonumber \ee
where $N_{\rm p}$ is the number or parameters and $N_{\rm b}$ is the number of $k$-bins used in the analysis.

\section{Results}

\begin{table*}
\begin{center}
\begin{tabular}{c c c c c c c c}
\hline\hline
Model & $\aone$ & $\az \ (\ab(z_{\rm eff}))$ &$\ap(z_{\rm eff})$ & $\alpha(z_{\rm eff})$ &  corr$(\az,\aone)$ & corr$(\ab(z_{\rm eff}),\ap(z_{\rm eff}))$ & $\chi^2 /{\rm DoF}$ \\ \hline
  & \multicolumn{7}{c}{Averaged mocks} \\
\hline
Fiducial&$-0.001\pm 0.252$& $1.000 \pm 0.086$&$0.999 \pm 0.213$&$1.000\pm 0.063$&$-0.585$&$-0.299$&$-$ \\
\hline
  & \multicolumn{7}{c}{DR14 QSO sample} \\
\hline
Fiducial & $-0.038 \pm 0.125$ &$1.037 \pm 0.059$&$0.998 \pm 0.097$& $1.024 \pm 0.040$ &$-0.723$&$-0.376$&$122/138$ \\
w/o $z$-weights &$-0.016\pm 0.130$&$1.027 \pm 0.069$&$1.011 \pm 0.090$&$1.022\pm0.048$&$-0.793$&$-0.419$&$56/66$\\
$k_{\rm max}=0.30 \hompc$ & $-0.026 \pm 0.118$ &$1.030 \pm 0.055$&$1.003 \pm 0.095$&$1.021\pm 0.038$&$-0.667$&$-0.284$&$164/194$ \\ 

 \hline\hline
\end{tabular}
\caption{Constraints on BAO parameters derived from the average of the EZ mocks (upper part of the table) and the DR14 quasar catalogue (lower). The analysis is performed with redshift weights using the $k$ modes in the range of $0.01<k<0.23 \ \hompc$. {To be imaginable, we also show $\alpha_{\perp}$, $\alpha_{\parallel}$ and their correlation at the effective redshift, where $\alpha_{\perp}$ is the same as $\alpha_{0}$.}}
\label{tab:BAO}
\end{center}
\end{table*}

We fit $\az$ and $\aone$ to the observables derived from both the EZ mocks and from the DR14 quasar sample using our template with redshift weights discussed in Sec. \ref{sec:temp}, setting $k_{\rm max}$ to $0.23 \hompc$ as the fiducial case. For comparison, we perform separate analyses in another two cases in which the redshift weights are not applied, or $k_{\rm max}$ is set to 0.3 $\hompc$.

The result of the mock test is listed in the upper section of Table \ref{tab:BAO}. The recovered values of the parameters from the mocks are in excellent agreement with the expected values of $\az=1, \aone=0$ in the fiducial case. We derive the constraints on $\ab, \ap$ and the isotropic BAO dilation parameter $\alpha\equiv\ab^{2/3}\ap^{1/3}$ from $\az$ and $\aone$, and quantify the correlation among these parameters. 

\begin{figure}
\centering
{\includegraphics[width=0.4\textwidth]{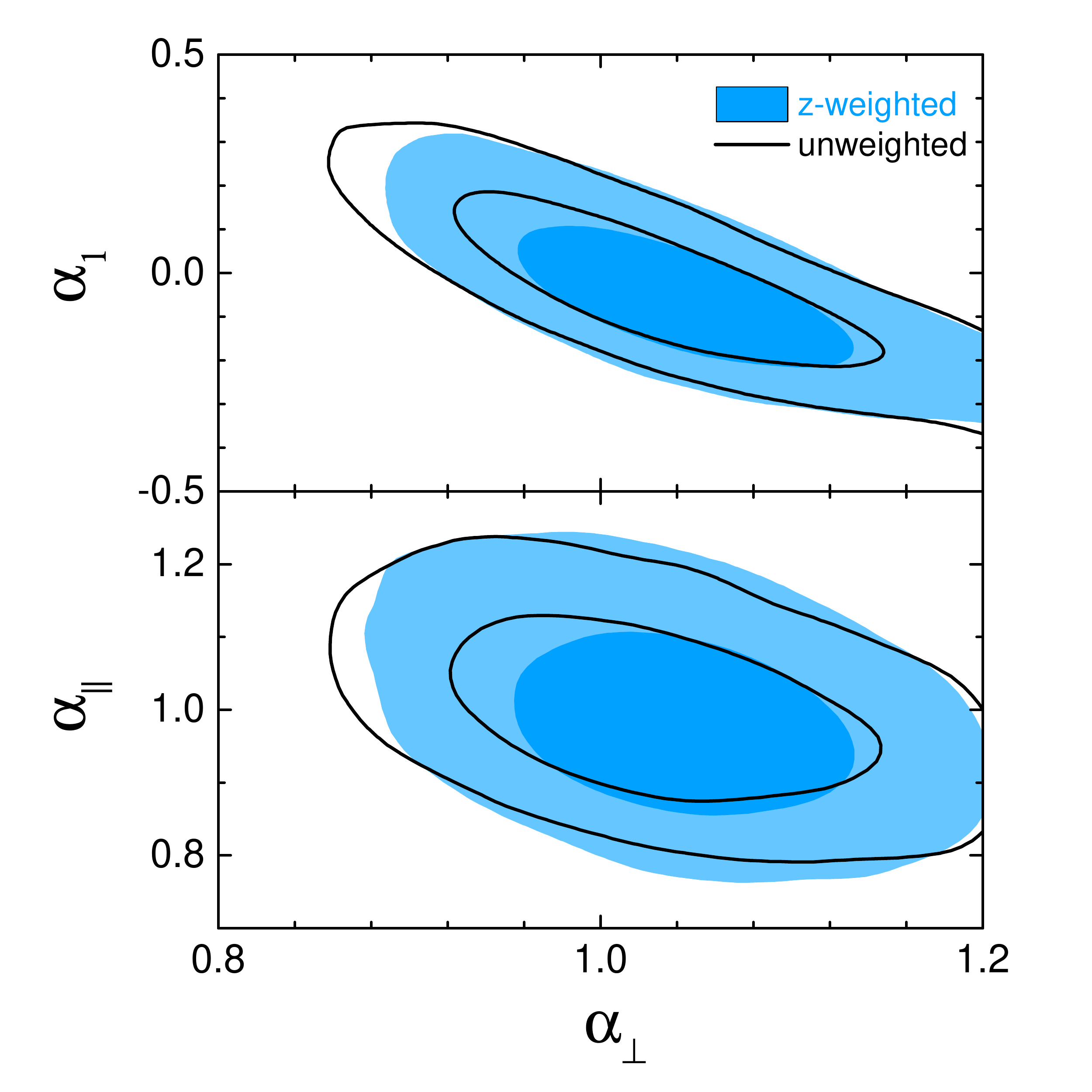}}
\caption{The 68 and 95\% CL contour plots for $\alpha_{0}(\ab(z_{\rm eff}))$, $\aone$ and $\ap(z_{\rm eff})$ with (blue filled) and without (black unfilled) the redshift weights. 
}
\label{fig:contour}
\end{figure}

\begin{figure}
\centering
{\includegraphics[scale=0.32]{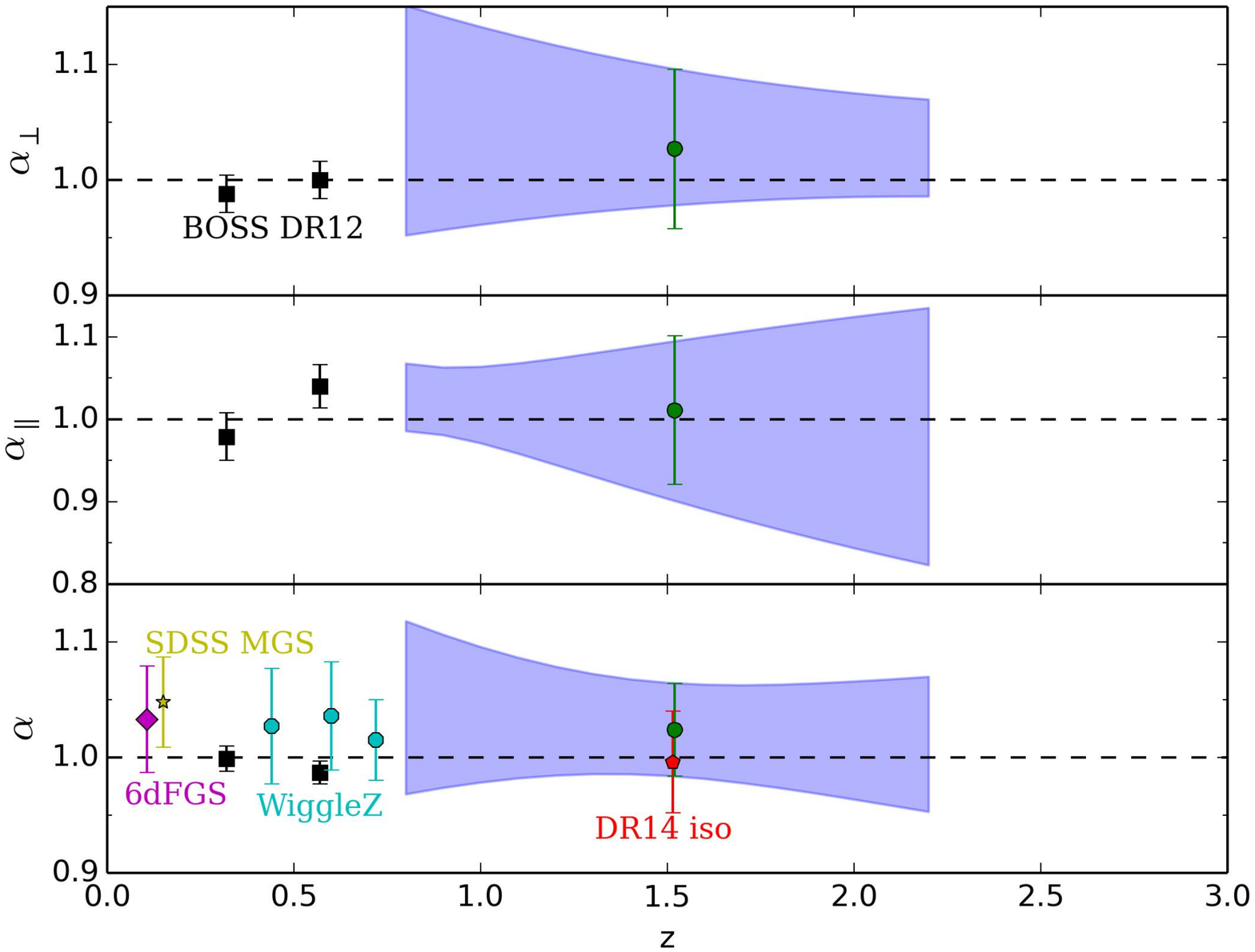}}
\caption{A 68\% CL reconstruction of the time evolution of $\ab$, $\ap$ and the isotropic $\alpha$ derived from from $\ab$, $\ap$ (blue bands), and the measurement at the effective redshift without $z$-weights (green circles with error bars), in comparison with other measurements in the literature including: BOSS DR12 \citep{Alam:2016hwk}, eBOSS  DR14  isotropic BAO constraint \citep{Ata:2017dya}, WiggleZ \citep{doi:10.1093/mnras/stu778} and MGS \citep{MGS}}
\label{fig:alpha}
\end{figure}

We then apply our pipeline to the DR14 quasar sample, and present the results in the lower part of Table \ref{tab:BAO} in cases of fiducial, no redshift weights, and with $k_{\max}$ extended to 0.3 $\hompc$. The results are illustrated in Figs \ref{fig:contour} and \ref{fig:alpha}. Our investigation reveals that,
\begin{itemize}
\item The redshift weights generally improve the constraints, with the uncertainty of $\az \ (\ab)$ and $\aone$ improved by $4\%$ and $14\%$, respectively. The uncertainty of the derived isotropic $\alpha$ are tightened by 17\%. The improvement is also visible from Fig \ref{fig:contour}, in which the 68 and 95\% confidence interval (CL) contours between $\ab,\ap$ and $\aone$ are shown; 
\item The constraint on the derived $\ap$ is not improved by the redshift weights. That is probably because the redshift weights reduce the correlation between $\az$ and $\aone$. This result explains why the constraint on $\ap$ can be diluted when both $\az$ and $\aone$ become better constrained ($\ap$ is essentially a product of $\az$ and $\aone$);
\item The parameter $\aone$ is consistent with zero within the error bars, that there is no evidence for the time evolution of $\alpha$ from the DR14 quasar sample. Fig \ref{fig:alpha}, shows a 68\% CL reconstruction of the evolution history of $\ab,\ap$ and $\alpha$ in the redshift range of $0.8<z<2.2$. For a comparison, we display the constraint without the redshift weights, along with other published constraints in the literature, including the BOSS result from \citep{Alam:2016hwk}, eBOSS  DR14  isotropic BAO constraint \citep{Ata:2017dya}, the WiggleZ \citep{doi:10.1093/mnras/stu778} and the MGS \citep{MGS} result;
\item Extending the $k$ range to 0.3 $\hompc$ slightly improves the constraints, i.e. the extension tightens the constraints on $\ab,\ap$ and $\alpha$ by $7\%$, $2\%$ and $5\%$, respectively, and the degeneracy among BAO parameters is slighted reduced. The constraint on $\alpha$,  $1.021\pm0.038$, is fully consistent with the BAO analysis using the same data sample \citep{Ata:2017dya};
\item The reduced $\chi^2$ in all cases is reasonably consistent with unity, meaning that we are neither overfitting or underfitting the data.
\end{itemize}

From our result, we derive the commonly used BAO distance indicators $D_{M}(r_{\rm d,fid}/r_{\rm d}), \ H(r_{\rm d}/r_{\rm d,fid})$ and $D_{V}(r_{\rm d,fid}/r_{\rm d})$ at five effective redshifts (Table \ref{tab:dist}), {and the correlations of $D_{M}(r_{\rm d,fid}/r_{\rm d})$ and $H(r_{\rm d}/r_{\rm d,fid})$ are shown in APPENDIX \ref{sec:handd}}. As these are derived from constraints on only two parameters of $\az$ and $\aone$, the error bars of these quantities are highly correlated, \ie, the covariance matrix is close to singular. Therefore these constraints are only suitable for comparison with other measurements at similar redshifts. For cosmological parameter estimation, we recommend the readers to use the result reported in Table \ref{tab:BAO} instead.

\begin{table}
\begin{center}
\begin{tabular}{c c c c}
\hline\hline

\multirow{2}{*}{Redshift}&
$D_{M}(r_{\rm d,fid}/r_{\rm d})$&$H(r_{\rm d}/r_{\rm d,fid})$&$D_{V}(r_{\rm d,fid}/r_{\rm d})$\\

 &$\rm[Mpc]$&$\rm[km s^{-1}Mpc^{-1}$]&$\rm[Mpc]$ \\
 \hline
 $z_{1}=0.8$&$3020\pm 258$&$104 \pm 4.4$& $2759\pm198$\\
 $z_{2}=1.0$&$3560\pm 292$&$118 \pm 5.6$& $3179\pm180$\\ 
 $z_{3}=1.5$&$4459\pm 275$&$159 \pm 15$& $3940\pm155$\\
 $z_{4}=2.0$&$5479\pm 239$&$207 \pm 29$& $4432\pm223$\\
 $z_{5}=2.2$&$5755\pm 235$&$227 \pm 35$& $4580\pm265$\\ 
 \hline \hline
\end{tabular}
\caption{The derived BAO distance indicators from Table \ref{tab:BAO}.}
\label{tab:dist}
\end{center}
\end{table}

\section{conclusion and discussions}

We have developed a method to extract the tomographic BAO information from wide-angle redshift surveys. Working in Fourier space, we analytically derive optimal redshift weights for power spectra multipoles for the eBOSS DR14 quasar sample, which covers the redshift range of $0.8<z<2.2$. We build a framework in which the redshift-weighted power spectra multipoles can be combined to yield improvement on the BAO constraint, and apply our pipeline to the DR14 quasar sample after validating it using the EZ galaxy mock catalogues. 

Our work yields an anisotropic BAO measurement at the effective redshift of $1.52$: $\ab=1.037\pm0.059$ and $\ap=0.998\pm0.097$, and an isotropic BAO measurement of $\alpha=1.024\pm0.040$. Compared to the case without the redshift weights, the constraint on the isotropic BAO dilation parameter gets tightened by $17\%$.

Another BAO analysis with redshift weights is performed in a companion paper \citep{Zhu18}, which differs from ours primarily regarding the fact that \cite{Zhu18} performs the analysis in configuration space. In this sense, our results are complementary to each other. The results from this work are generally consistent with that in \cite{Zhu18}.

Two additional companion papers \citep{Zhao18,RR18} perform joint BAO and RSD analysis with the optimal redshift weights in Fourier space. Our BAO constraints are generally consistent with each other within the error budget. 

The method developed in this work can be directly applied to the complete eBOSS sample when the survey finishes, and to future deep redshift surveys including DESI \citep{DESI} and Euclid \citep{Euclid}.

\section*{Acknowledgements}

DW, GB and YW are supported by NSFC Grants 1171001024 and 11673025. GBZ is also supported by a Royal Society Newton Advanced Fellowship, hosted by University of Portsmouth. YW is supported by a NSFC Grant No. 11403034, and by a Young Researcher Grant of National Astronomical Observatories, Chinese Academy of Sciences.

Funding for SDSS-III and SDSS-IV has been provided by
the Alfred P. Sloan Foundation and Participating Institutions.
Additional funding for SDSS-III comes from the
National Science Foundation and the U.S. Department of Energy Office of Science. Further information about
both projects is available at \url{www.sdss.org}.
SDSS is managed by the Astrophysical Research Consortium
for the Participating Institutions in both collaborations.
In SDSS-III these include the University of
Arizona, the Brazilian Participation Group, Brookhaven
National Laboratory, Carnegie Mellon University, University
of Florida, the French Participation Group, the German Participation Group, Harvard University,
the Instituto de Astrofisica de Canarias, the Michigan
State / Notre Dame / JINA Participation Group, Johns
Hopkins University, Lawrence Berkeley National Laboratory,
Max Planck Institute for Astrophysics, Max Planck
Institute for Extraterrestrial Physics, New Mexico State
University, New York University, Ohio State University,
Pennsylvania State University, University of Portsmouth,
Princeton University, the Spanish Participation Group,
University of Tokyo, University of Utah, Vanderbilt University,
University of Virginia, University of Washington,
and Yale University.

The Participating Institutions in SDSS-IV are
Carnegie Mellon University, Colorado University, Boulder,
Harvard-Smithsonian Center for Astrophysics Participation
Group, Johns Hopkins University, Kavli Institute
for the Physics and Mathematics of the Universe
Max-Planck-Institut fuer Astrophysik (MPA Garching),
Max-Planck-Institut fuer Extraterrestrische Physik
(MPE), Max-Planck-Institut fuer Astronomie (MPIA
Heidelberg), National Astronomical Observatories of
China, New Mexico State University, New York University,
The Ohio State University, Penn State University,
Shanghai Astronomical Observatory, United Kingdom
Participation Group, University of Portsmouth, University
of Utah, University of Wisconsin, and Yale University.

This work made use of the facilities and staff of the UK Sciama High Performance Computing cluster supported by the ICG, SEPNet and the University of Portsmouth.
This research used resources of the National Energy Research
Scientific Computing Center, a DOE Office of Science User Facility 
supported by the Office of Science of the U.S. Department of Energy 
under Contract No. DE-AC02-05CH11231.

\appendix
\section{THE CORRELATION MATRIX}
\label{sec:handd}
\begin{multline} \label{5bins} \left[\begin{array}{cccccc}
 D_{M}(z_{1})&H(z_{1})&D_{M}(z_{2})&H(z_{2})&D_{M}(z_{3})&H(z_{3})\\
 \\
1.000&    0.565&    0.997&   -0.106&   0.948&   -0.657\\ 
 &    1.000&    0.623&    0.761&    0.798&    0.252\\ 
 &    &    1.000&   -0.034&    0.969&   -0.600\\
  &    &   &    1.000&    0.216&    0.820\\ 
 &    &    &    &    1.000&   -0.382\\
 &   &   &  &   &    1.000\\
  &  & & & & \\ 
& & & & & \\ 
&&&&&\\
&&&&&\\
\end{array}\right.\\
\left.\begin{array}{cccccc}
D_{M}(z_{4}))&H(z_{4})&D_{M}(z_{5})&H(z_{5})\\
\\
0.767&   -0.760&    0.643&   -0.779 && D_{M}(z_{1})\\
0.963&    0.108&    0.995&    0.781&&H(z_{1})\\
0.812&   -0.710&    0.697&   -0.731&&D_{M}(z_{2})\\
0.556&    0.728&    0.693&    0.707&&H(z_{2})\\
0.931&   -0.513&    0.853&   -0.539&& D_{M}(z_{3})\\
-0.020&    0.989&    0.155&    0.985&&H(z_{3})\\
1.000&   -0.165&    0.985&   -0.195&& D_{M}(z_{4})\\
 & 1.000&    0.010&    1.000&&H(z_{4})\\
  &&1.000&   -0.020&& D_{M}(z_{5})\\
  &&&   1.000&&H(z_{5})\\
\end{array}\right]
\end{multline}
The correlation matrix of Fig \ref{tab:dist}.

\bibliographystyle{mn2e}
\bibliography{paperweight}

\end{document}